\documentclass[a4paper, 11pt]{article}
\pdfoutput=1
\usepackage{jheppub}
\usepackage[colorlinks = true, citecolor = blue, linkcolor = blue, urlcolor = blue]{hyperref}

\usepackage[utf8]{inputenc}

\usepackage{xcolor}

\usepackage{amsmath}
\usepackage{amssymb}
\usepackage{tikz}
\usepackage{siunitx}
\usepackage{physics}
\usepackage{slashed}

\usepackage{overpic}

\usepackage{graphicx}

\usepackage{mathtools}

\newcommand{\rt}{{\mathbf{r}}}
\newcommand{\xt}{{\mathbf{r}}}

\newcommand{\bt}{{\mathbf{b}}}

\newcommand{\nc}{N_c}

\newcommand{\aem}{\alpha_\text{em}}

\newcommand{\LambdaQCD}{\Lambda_\text{QCD}}

\usepackage{graphicx}
\graphicspath{{figures/}}
\usepackage{multirow}
\usepackage{tabularx}
\usepackage{physics}
\usepackage[normalem]{ulem}
\usetikzlibrary{shapes.symbols}
\usepackage{tikz}
\usepackage{mathtools}
\usepackage[capitalise]{cleveref} 
\usepackage{bm}
\usepackage{mathrsfs}
\usepackage{xcolor}
\usepackage{diagbox}
\usepackage{subfigure}
\definecolor{tangerine}{RGB}{242, 133, 0}
\definecolor{persianGreen}{RGB}{0, 166, 147}
\usepackage{booktabs}
\usepackage{adjustbox}
\usepackage{subcaption}

\makeatletter
\def\@fpheader{~}
\makeatother

\title{Determination of the initial condition for the Balitsky-Kovchegov equation with transformers}

\author[a,b,h]{Meisen Gao,}
\author[c,d,e]{Zhong-Bo Kang,}
\author[c,d]{Jani Penttala,}
\author[b,f,g]{and Ding Yu Shao}

\affiliation[a]{School of Physics, East China University of Science and Technology, Shanghai 200237, China}
\affiliation[b]{Department of Physics and Center for Field Theory and Particle Physics, Fudan University, Shanghai, 200433, China}
\affiliation[c]{Department of Physics and Astronomy, University of California, Los Angeles, CA 90095, USA}
\affiliation[d]{Mani L. Bhaumik Institute for Theoretical Physics, University of California, Los Angeles, CA 90095, USA}
\affiliation[e]{Center for Frontiers in Nuclear Science, Stony Brook University, Stony Brook, NY 11794, USA}
\affiliation[f]{Key Laboratory of Nuclear Physics and Ion-beam Application (MOE), Fudan University, Shanghai, 200433, China}
\affiliation[g]{Shanghai Research Center for Theoretical Nuclear Physics, NSFC and Fudan University, Shanghai 200438, China}
\affiliation[h]{Shanghai Key Laboratory of Particle
Physics and Cosmology, Shanghai 200240, China}

\emailAdd{msgao@ecust.edu.cn}
\emailAdd{zkang@physics.ucla.edu}
\emailAdd{janipenttala@physics.ucla.edu}
\emailAdd{dingyu.shao@cern.ch}

\abstract{
In the high-energy limit of QCD, scattering off nucleons and nuclei can be described in terms of Wilson-line correlators whose energy dependence is perturbative. The energy dependence of the two-point correlator, called the dipole amplitude, is governed by the Balitsky--Kovchegov (BK) equation. The initial condition for the BK equation can be fitted to the experimental data, which requires evolving the dipole amplitude for a large set of different parameter values. In this work, we train a transformer model to learn the energy dependence of the dipole amplitude, thereby replacing repeated time-consuming numerical BK evolutions during the fit by fast emulator predictions. The transformer predicts the learned dipole amplitude and the leading order inclusive deep inelastic scattering cross section very accurately, allowing for efficient fitting of the initial condition to the experimental data. Using this setup, we fit the initial condition of the BK equation to the inclusive deep inelastic scattering data from HERA and consider two different starting points $x_0$ for the evolution. We find better agreement with the experimental data for a smaller $x_0$. This work paves the way for future studies involving global fits of the dipole amplitude at leading order and beyond.
}

\begin{document}

\maketitle

\section{Introduction} \label{sec:intro}

Mapping the partonic structure of protons and nuclei in the high-energy regime is a foremost challenge that provides the primary scientific motivation for the future Electron-Ion Colliders~\cite{Accardi:2012qut, AbdulKhalek:2021gbh,Anderle:2021wcy}. At a very small Bjorken-$x$ variable, the density of gluons within a hadron grows rapidly, leading to the gluon saturation phenomenon where non-linear QCD dynamics become dominant. The Color Glass Condensate (CGC) effective field theory~\cite{Iancu:2003xm, Gelis:2010nm} provides a powerful framework for describing this dense gluonic matter. Central to the CGC is the dipole--target scattering amplitude, which encodes the essential information about the nonperturbative scattering off the gluonic target. The energy evolution of the dipole amplitude is governed by perturbative evolution equations, most notably the Balitsky--Kovchegov (BK) equation~\cite{Balitsky:1995ub, Kovchegov:1999yj}. However, the starting point for this evolution---the initial condition at a moderately small $x_0$---is fundamentally nonperturbative and must be determined from experimental data.

A convenient initial condition for the high-energy evolution is given by the McLerran--Venugopalan (MV) model~\cite{McLerran:1993ka, McLerran:1993ni, McLerran:1994vd}, which allows for a description of the dipole amplitude in terms of a few free parameter that can be extracted from the experimental data.
The MV model, and its generalizations, have then be used in many successful extractions at both leading order~\cite{Albacete:2009fh, Lappi:2013zma, Albacete:2010sy, Ducloue:2019jmy, Beuf:2020dxl, Dumitru:2023sjd, Casuga:2023dcf, Hanninen:2025iuv} and next-to-leading order~\cite{Hanninen:2022gje, Casuga:2025etc} using the inclusive deep inelastic scattering (DIS) data from HERA~\cite{H1:2009pze, H1:2015ubc}.
The extracted parameters contain the fundamental information about the target structure, including the saturation scale $Q_s$ that describes the onset of the gluon saturation phenomenon.

However, to determine the initial condition from the experimental data, one has to run the BK evolution for a wide range of parameter values.
Due to the nonlinear nature of the BK evolution, its evaluation is much more demanding than the linear BFKL~\cite{Lipatov:1976zz, Fadin:1975cb, Kuraev:1976ge, Kuraev:1977fs, Balitsky:1978ic}
and DGLAP~\cite{Dokshitzer:1977sg, Gribov:1972ri, Gribov:1972rt, Altarelli:1977zs} evolution equations, making the BK evolution the bottleneck for numerical studies even at leading order.
At higher orders in perturbation theory, the cross section itself becomes also numerically demanding due to the large amount of integrals over transverse coordinates that are typical for the dipole picture used in small-$x$ calculations.
For these reasons, it becomes important to study efficient methods for evaluating these quantities with different models for the dipole amplitude.

To overcome this limitation, we turn to the rapidly advancing field of machine learning and neural-network emulation.
We employ the \textit{transformer} architecture, originally introduced for sequence modelling with self-attention~\cite{Vaswani:2017Attention}.
Transformers are widely used to capture long-range dependencies in sequential data, including long-horizon time-series forecasting~\cite{Zhou:2021Informer,Lim:2021TFT}.
In high-energy physics, transformer architectures have also been successfully applied to learn correlations in variable-length particle sets, e.g., for jet tagging~\cite{Qu:2022ParT} and symmetry-aware event reconstruction tasks such as jet--parton assignment~\cite{Shmakov:2022SPANet,Fenton:2022Permutationless}.

Motivated by these developments, we treat the model parameters together with the kinematic variables as an ordered sequence of inputs.
In this way, the transformer's self-attention mechanism~\cite{Vaswani:2017Attention} can learn intricate, non-local correlations in the BK-evolved dipole amplitude and the corresponding DIS dataset (such as those from HERA) without imposing a rigid functional ansatz.

From a functional viewpoint, the BK evolution is defined as a non-linear differential equation that resums logarithms of energy. For a given set of
initial-condition parameters and a given kinematic point $(r,x)$, it returns the dipole amplitude
$N(r,x)$ that has been evolved starting from an initial condition at $x_0 > x$.
In practical fits one
must evaluate this map for many parameter choices and across a wide kinematic domain spanning many
orders of magnitude in both $r$ and $x$. As a result, the relevant target function is strongly
multi-scale and exhibits correlated variations between the initial-condition parameters and the
kinematic variables.

The self-attention mechanism in transformers provides a flexible way to model such cross-dependencies:
by allowing each input component (parameters and kinematics) to interact with all others, the network
can learn global correlations that are difficult to capture with rigid low-dimensional ansätze.
This perspective is closely related to recent developments in scientific machine learning, where
attention-based architectures have been explored for surrogate/operator learning of complex non-linear
systems~\cite{Hao:2023GNOT,Li:2023FactFormer}.

To our knowledge, this paper presents the first determination of the small-$x$ dipole amplitude using this transformer-based emulator strategy.
Moreover, we consider different starting points for the BK evolution, enabling us to study potential biases in the fitted initial condition and to assess the applicability of BK evolution when starting from more moderate values of Bjorken $x$.
It should be emphasized that the present analysis is restricted to leading-order accuracy, and our primary aim is to demonstrate the methodology rather than to obtain the most precise phenomenological fit.

An alternative machine-learning approach based on
Gaussian-process emulators has been successfully applied in small-$x$ studies before~\cite{Casuga:2023dcf,Casuga:2025etc},
providing flexible surrogate models with quantified uncertainties.
In this work, we employ a transformer-based approach, which can naturally accommodate larger training sets,
capture correlations across the parameter space, and demonstrates stable performance for learning such
highly non-linear multi-variable mappings~\cite{Hao:2023GNOT,Li:2023FactFormer}.
This makes it well suited for global analyses requiring high precision over a wide kinematic range.

The paper is organized as follows. In Sec.~\ref{sec:theory}, we briefly review the theoretical framework, including the dipole formalism in DIS and the BK equation. In Sec.~\ref{sec:fitting}, we detail our neural network architecture, the training methodology, and how we incorporate necessary physical constraints. In Sec.~\ref{sec:application_dis}, we present the resulting dipole amplitude and perform a rigorous comparison against both the HERA data and the results from traditional parameterization-based fits. Finally, in Sec.~\ref{sec:discussion}, we summarize our conclusions and provide an outlook on how this data-driven approach can pave the way for precision studies at the future EIC.

\section{Theoretical framework}
\label{sec:theory}

\subsection{Inclusive DIS in the dipole picture}

At small $Q^2$ region, where contributions from $Z$ boson exchange are neglected, inclusive DIS is described by the structure functions $F_2(x, Q^2)$ and $F_L(x, Q^2)$, which are related to the experimental reduced cross section $\sigma_r$ by
\begin{equation}
    \sigma_r(y,x,Q^2) = F_2(x,Q^2) - \frac{y^2}{1+ (1-y)^2} F_L(x, Q^2),
\end{equation}
with
\begin{align}
    F_2(x,Q^2) = F_L(x,Q^2) + F_T(x,Q^2), 
\end{align}
where $Q^2$ is the photon virtuality, $x$ is the Bjorken scaling variable, and $y$ is the inelasticity. The structure functions are defined in terms of the cross sections for the scattering of a transversely ($T$) or longitudinally ($L$) polarized virtual photon off a proton target
\begin{align}
    F_\lambda(x,Q^2) \equiv \frac{Q^2}{4\pi^2 \aem} \sigma^{\gamma^*_\lambda p},
\end{align}
where $\aem$ is the fine-structure constant and $\lambda=T$ or $L$ denotes the polarization of the virtual photon.

In the small-$x$ regime, these cross sections can be calculated within the dipole picture~\cite{Kovchegov:2012mbw}, where the interaction factorizes into the splitting of the virtual photon into a quark--antiquark pair ($q\bar{q}$), followed by the scattering of this color dipole off the proton target. The total cross section is given by an integral over the dipole's transverse size $\mathbf{r}$, its impact parameter $\mathbf{b}$, and the quark's longitudinal momentum fraction $z$, which reads
\begin{equation}\label{eq:cross_section}
     \sigma^{\gamma^*_\lambda p} = \frac{4 \aem \nc}{(2\pi)^2} \sum_f e_f^2
    \int \dd[2]{\rt} \dd[2]{\bt} \int_0^1 \dd{z} \mathcal{K}_\lambda(\rt,z) N(\rt, \bt, x),
\end{equation}
where the sum runs over all active quark flavors $f$. The dynamics of the strong interaction are encoded in the nonperturbative dipole amplitude $N(\mathbf{r}, \mathbf{b}, x)$, which describes the probability for the dipole to scatter. The perturbative component is contained in the squared photon light-front wave functions, $\mathcal{K}_\lambda$, which are given by
\begin{align}
    \mathcal{K}_L &= 4 \overline Q^2 z(1-z) K_0 \qty( \overline Q \abs{\rt} )^2, \\
    \mathcal{K}_T &= \overline Q^2  \qty[z^2+(1-z)^2] K_1 \qty( \overline Q \abs{\rt} )^2 ,
\end{align}
and the argument $\overline Q$ of the modified Bessel functions $K_{0}$ and $K_1$ is $\overline Q^2 \equiv z(1-z) Q^2$. Here, we have ignored quark mass corrections.
In many dipole-model phenomenological studies, quark-mass threshold effects are approximately accounted for by a rescaling of the Bjorken variable,
$\tilde{x}=x\,(1+4m_f^2/Q^2)$ (the original GBW prescription)~\cite{Golec-Biernat:1998zce}.
In this work, since we adopt massless photon wave functions at leading order, we do not apply this $x$-rescaling.

In Eq.~\eqref{eq:cross_section}, the full dipole--proton scattering amplitude $N(\rt, \bt, x)$ depends on the transverse size of the dipole $\rt$, Bjorken $x$, and the impact parameter $\bt$. However, inclusive DIS data depends only on the dipole amplitude integrated over the impact parameter. We therefore simplify the calculation by integrating out the impact parameter as
\begin{equation}
    \int \dd[2]{\bt} N(\rt, \bt ,x) \equiv \frac{\sigma_0}{2} N(r , x)\,,
\end{equation}
and introducing a single normalization parameter, $\sigma_0/2$, which has the interpretation as the \textit{proton transverse area} ~\cite{Lappi:2013zma, Albacete:2010sy, Albacete:2009fh}. 
This parameter effectively sets the total strength of the interaction by replacing the impact parameter integral in the cross-section calculation. It is a free parameter in our model, constrained by the fit to HERA data.
The integrated dipole amplitude $N(r, x)$ now only depends on the dipole size $r = \abs{\rt}$ and the Bjorken $x$ variable.
Due to rotational symmetry, there is no dependence on the angle of the vector $\rt$ after integrating over the impact parameter.

\subsection{Balitsky--Kovchegov evolution}

The energy dependence of the dipole amplitude, $N(r, x)$, is governed by the BK evolution equation. This equation resums quantum corrections proportional to large logarithms of energy, 
typically written in terms of the rapidity variable $Y=\ln(1/x)$, and incorporates the non-linear effects that lead to gluon saturation. For a dipole with quark and antiquark at transverse positions $\xt_0$ and $\xt_1$, the equation reads
\begin{multline}\label{eq:BK_eq}
    \frac{\partial N(\abs{\xt_{01}},x)}{\partial Y} = \int \mathrm{d}^2\xt_2 \, K_{\text{BK}}(\xt_0, \xt_1, \xt_2) \\
    \times \left[ N(\abs{\xt_{02}},x) + N(\abs{\xt_{12}},x) - N(\abs{\xt_{01}},x) - N(\abs{\xt_{02}},x)N(\abs{\xt_{12}},x) \right],
\end{multline}
where $\xt_{ij} \equiv \xt_i - \xt_j$ and we have ignored the impact-parameter dependence. The linear terms in the square brackets correspond to the BFKL equation, describing the emission of a gluon from the parent dipole. The final quadratic term, $N(\abs{\xt_{02}},x)N(\abs{\xt_{12}},x)$, represents the simultaneous scattering of the two new dipoles, a non-linear effect that tames the growth of the amplitude and drives the system towards saturation.
We note that there also additional corrections to the BK equation that incorporate the correct time-ordering of the emitted gluons~\cite{Beuf:2014uia,Iancu:2015vea,Ducloue:2019ezk,Boussarie:2025mzh,Boussarie:2025bpq}, which would be important at next-to-leading order~\cite{Lappi:2016fmu,Ducloue:2019ezk}.

The interaction kernel, $K_{\text{BK}}$, describes the splitting of the parent dipole. For phenomenological accuracy, it is crucial to include running coupling corrections. Following the Balitsky prescription~\cite{Balitsky:2006wa}, the kernel is given by
\begin{multline}
    K_{\text{BK}}(\xt_0, \xt_1, \xt_2) = \frac{N_c \, \alpha_s(\xt_{01}^2)}{2\pi^2} \left[ \frac{\xt_{01}^2}{\xt_{02}^2 \xt_{12}^2}
    + \frac{1}{\xt_{02}^2}\left(\frac{\alpha_s(\xt_{02}^2)}{\alpha_s(\xt_{12}^2)} - 1\right) + \frac{1}{\xt_{12}^2}\left(\frac{\alpha_s(\xt_{12}^2)}{\alpha_s(\xt_{02}^2)} - 1\right) \right].
\end{multline}
Here the position-space strong coupling constant, $\alpha_s(r^2)$, is taken at leading order with $n_f=3$ active quark flavors as~\cite{Lappi:2013zma}
\begin{equation}
    \alpha_s(r^2) = \frac{12\pi}{(33-2n_f) \ln\left( \frac{4C^2}{r^2 \LambdaQCD^2} \right)}.
\end{equation}
Here, $\LambdaQCD = \SI{0.241}{GeV}$ and $C^2$ is a parameter that sets the scale of the argument of the logarithm, which must be determined from fits to data \cite{Lappi:2013zma, Casuga:2023dcf}. For large values of the dipole size $r$, we freeze $\alpha_s$ to the value $\alpha_{s,\text{max}}=0.7$.

\section{Fitting procedure}
\label{sec:fitting}

The parameters of the initial dipole scattering amplitude are determined from deep-inelastic scattering data via a two-stage procedure. First, the BK evolution equation is solved for a large ensemble of initial conditions to generate a comprehensive library of theoretical predictions. Second, this library is used to train a neural network emulator that interpolates the solutions of the BK evolution across the parameter space. The resulting high-quality model allows for the rapid and efficient execution of a global fit to HERA data within the dipole picture.

\subsection{Sampling and solving the BK equation}

The BK equation $\eqref{eq:BK_eq}$ describes the evolution of the dipole amplitude $N(r, x)$ with respect to the collision energy.
The evolution is uniquely specified by an initial condition at $x_0$, and to assess the dependence on the initial rapidity scale we consider two different choices $x_0 =0.01$ and $x_0=0.05$.
For the initial condition, we use a generalization of the MV model considered in Ref.~\cite{Lappi:2013zma}
\begin{equation}
    N(r,x_0) = 1 - \exp\bigg[-\biggl(\frac{r^2 Q_{s0}^2}{4}\biggr)^\gamma \ln\Bigl(\frac{1}{\LambdaQCD\,r} + e\,e_c\Bigr)\bigg],
\end{equation}
which is a function of the initial saturation scale $Q_{s0}$, the anomalous dimension $\gamma$, and an infrared regulator $e_c$. To efficiently map this parameter space, we use Latin hypercube sampling (LHS) to generate $\num{10000}$ parameter sets for $(Q_{s0},\gamma,e_c,C^2)$. For each set, the running-coupling BK equation with the Balitsky kernel is solved using a numerical Julia implementation developed for this work. 
The solution is tabulated on a two-dimensional grid spanning transverse sizes $10^{-6}\leq r/\mathrm{GeV}^{-1}\leq 10^{2}$ and rapidities $0\leq \ln(x_0/x) \leq 16$, which covers the relevant HERA kinematics~\cite{H1:2015ubc}. This procedure yields a dataset of approximately $4.7\times 10^7$ values of $N(r,x)$, which serves as the training data for the neural network model.

For the neural-network inputs and diagnostics, we work in log-transformed variables for the two scale parameters,
using $\ln Q_{s0}$ and $\ln C^2$, which improves numerical conditioning across the wide dynamic ranges.
The size of the training library (10\,000 configurations) is chosen as a practical balance between the dimensionality
of the parameter space, the target emulator accuracy, and the computational cost of generating BK solutions.
In general, using substantially fewer configurations reduces the coverage of the parameter space and tends to increase
the emulator errors, while increasing the library size improves accuracy at the expense of additional BK-evolution time.

The BK solutions are tabulated as $N(r,x)$ on a two-dimensional output grid summarized in
Table~\ref{tab:bk_grid}. The $r$ grid is logarithmic (uniform in $\ln r$), while the $x$ output
points are uniformly spaced in $\ln(x_0/x)$ with step $\Delta\ln(x_0/x)=0.2$.
Here $\Delta\ln(x_0/x)$ denotes only the output spacing.
To reduce the size of the training dataset, we keep only samples with $\ln N(r,x)\ge -10$ when
forming the training set; this cut is applied only in data preprocessing and does not affect the BK
evolution itself.
\begin{table}[t]
\centering
\begin{tabular}{l c}
\toprule
Quantity & Value \\
\midrule
$r$ grid & logarithmic (uniform in $\ln r$) \\
$r_{\min}$ [$\mathrm{GeV}^{-1}$] & $10^{-6}$ \\
$r_{\max}$ [$\mathrm{GeV}^{-1}$] & $10^{2}$ \\
$N_r$ & $150$ \\
\midrule
$x$ output grid & uniform in $\ln(x_0/x)$ \\
$\ln(x_0/x)_{\min}$ & $0$ \\
$\ln(x_0/x)_{\max}$ & $16$ \\
$\Delta\ln(x_0/x)$ & $0.2$ \\
\bottomrule
\end{tabular}
\caption{Output grid used to store the BK solutions. The $x$ points are logarithmically spaced,
corresponding to uniform spacing in $\ln(x_0/x)\in[0,16]$ with $\Delta\ln(x_0/x)=0.2$.}
\label{tab:bk_grid}
\end{table}

\subsection{Neural network emulation of the BK amplitude}

\begin{figure}[t]
  \centering
  \includegraphics[width=\textwidth]{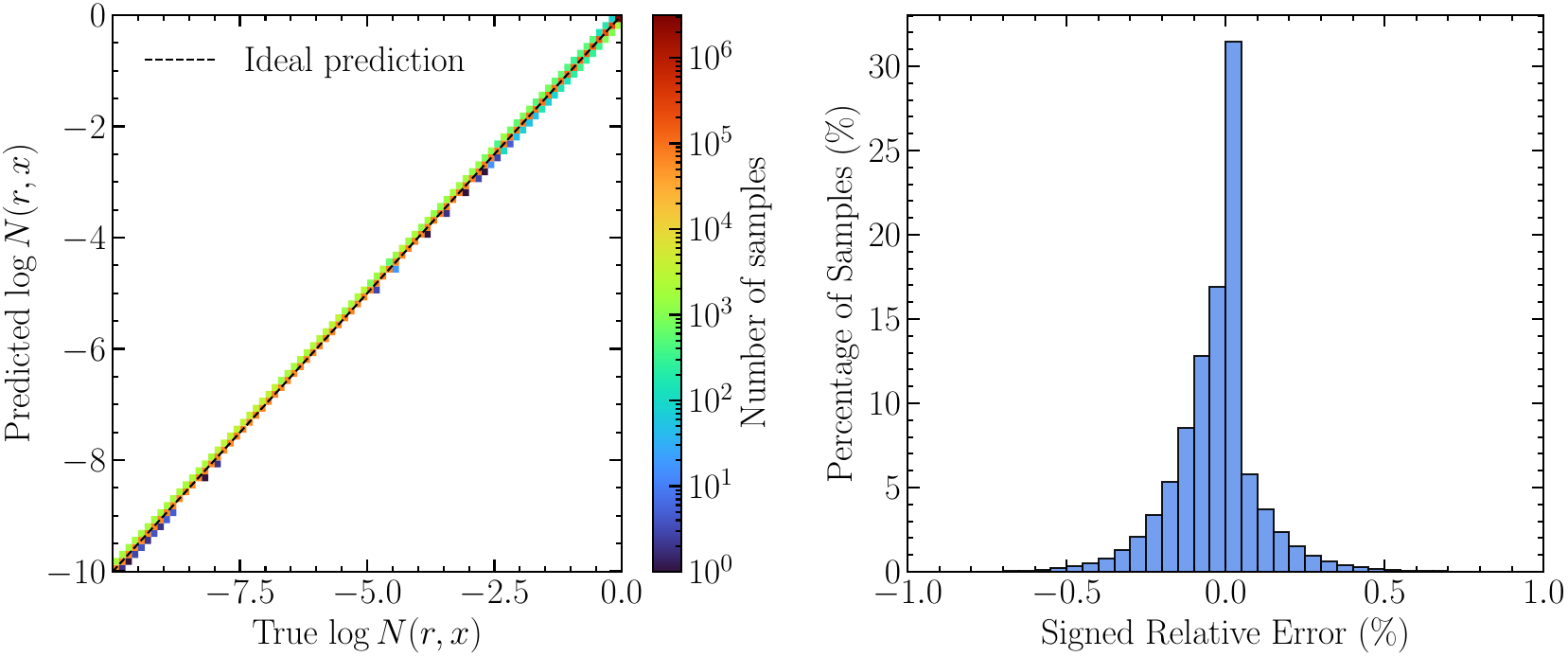}
  \caption{Left: Two dimensional histogram comparing the exact and transformer emulated values of $\ln N(r,x)$ for the BK dipole amplitude. The color scale indicates the density of samples, and the dashed line represents exact agreement. The mean and median relative errors on the validation set are $0.09\%$ and $0.05\%$, respectively.
  Right: Distribution of the signed relative errors, zoomed to $\pm1\%$.
  This highlights that the vast majority of predictions fall within a few per mille of the true values,
  demonstrating the emulator's high accuracy. We find that $99.978\%$ of the validation points fall within
  the $\pm1\%$ interval.}
  \label{fig:bk_learning}
\end{figure}

Directly solving the BK equation repeatedly across the parameter space would be computationally expensive. To circumvent this we train a neural network to emulate the functional dependence of $\ln N(r, x)$ on the variables $(r,x)$ and the four parameters of the BK setup. Owing to its ability to capture correlations in an ordered sequence of input variables, we adopt a transformer architecture.
The network takes as input the 6-dimensional vector
$(\ln Q_{s0},\,\allowbreak e_c,\,\allowbreak \gamma,\,\allowbreak
\ln C^2,\,\allowbreak \ln (r/\mathrm{GeV}^{-1}),\,\allowbreak \ln (x_0/x))$.
A single input vector corresponds to one training sample, i.e. one tabulated BK value at a specific grid point $(r,x)$
for a specific parameter set $\boldsymbol{\theta}=(Q_{s0},e_c,\gamma,C^2)$. Concretely, the first four entries specify the BK setup
(the initial-condition parameters $\boldsymbol{\theta}$), and the last two entries specify the kinematics at which the amplitude is evaluated.
The full training dataset is constructed by looping over all sampled parameter sets and, for each of them, over all $(r,x)$ points on the
numerical output grid used to tabulate the BK solutions.
Each scalar is mapped by a shared linear projection to form a sequence of six tokens of dimension $d_{\rm model}=128$ (see App.~\ref{app:transformer_nomenclature} for terminology). A sinusoidal positional encoding is added to the sequence, which is then processed by a transformer encoder with four layers, eight attention heads, feed-forward width~512, dropout~0.1, and pre–normalization.  The token representations are mean-pooled and passed to a two-layer multilayer perceptron (MLP) head \cite{GARDNER19982627} that outputs a raw logit function \cite{pmlr-v162-wei22d}.  
Applying a sigmoid to this logit yields the dipole amplitude $N(r, x)$, automatically enforcing the unitarity bound $0\leq N \leq 1$.  For diagnostics we also use $\log N$ after the sigmoid.

For training we adopt a “log-MSE” loss: the network’s raw logit is passed through a sigmoid to yield \(N_{\rm pred}\in(0,1)\), and we minimise the mean squared error between \(\log N_{\rm pred}\) and \(\log N_{\rm true}\). This formulation with $\log N$ simultaneously accommodates the dilute regime (\(N\ll1\)) and the saturation regime (\(N\approx1\)) without the need for a piece-wise or hybrid design. 

Inputs are standardized using the \textsc{StandardScaler} implementation from the Scikit-learn library~\cite{pedregosa2011scikit}. The model is trained using the AdamW optimizer \cite{Loshchilov2017DecoupledWD} with learning rate $3\times10^{-4}$, weight decay $10^{-4}$, cosine annealing schedule to $3\times10^{-6}$, batch size~$4096$, and early stopping (patience~$20$).
Automatic
mixed-precision training~\cite{Micikevicius2017MixedPT} and gradient clipping (max-norm~$1.0$) are applied. $90\%$ of the generated data is used for training and the remainder for validation.

\begin{figure}[t]
  \centering
  \includegraphics[width=\linewidth]{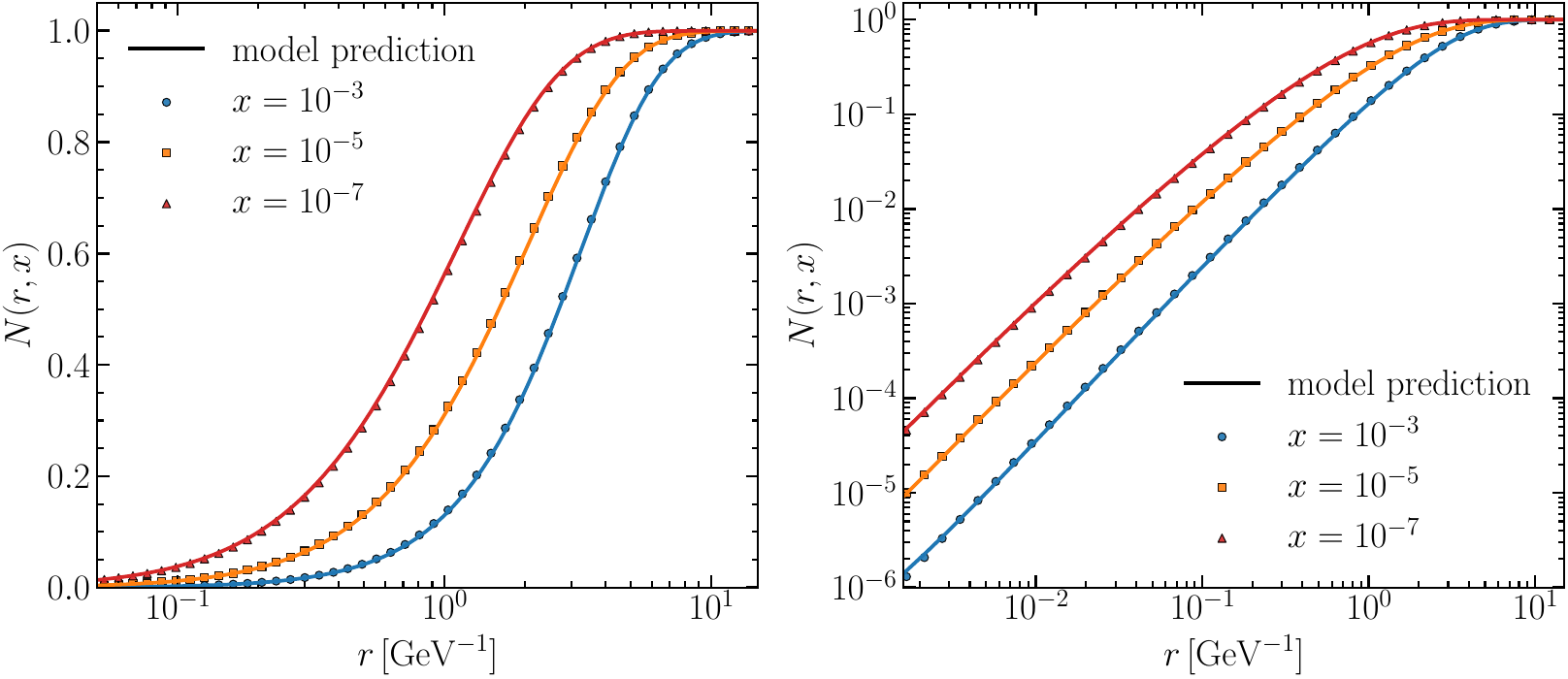}
  \caption{The emulated dipole amplitude $N(r,x)$ (lines) is compared against the exact BK solution (points) for a representative out-of-training-sample parameter set at three different rapidities, $x = 10^{-3}$, $10^{-5}$, and $10^{-7}$. The excellent agreement in both log-linear (left) and log-log (right) scales across the entire range of $r$ demonstrates the model's strong generalization performance on a held-out parameter set. The chosen parameters are $Q_{s0}^2=0.07$\,GeV$^2$, $\gamma=1.01$, $e_c=24.68$, $C^2=4.65$, and $x_0=0.01$, corresponding roughly to physical values found in previous fits~\cite{Lappi:2013zma,Casuga:2023dcf}.}
  \label{fig:N_vs_r}
\end{figure}

The trained emulator exhibits excellent performance, achieving mean and median relative errors of $0.09\%$ and $0.05\%$, respectively, on the validation set. This high accuracy is visually demonstrated in Fig.~\ref{fig:bk_learning}. The left panel shows a strong one-to-one correlation between the emulated and exact values of $\ln N(r, x)$ across many orders of magnitude. The distribution of relative errors, shown in the right panel, is sharply peaked at zero, confirming that the vast majority of predictions are accurate to within a few per mille. The slight asymmetry in the 2D histogram reflects the non-uniform distribution of validation
samples across the multi-scale $(r,x)$ domain and parameter space, and is amplified by binning in the density plot.
The relative errors quantify the emulator accuracy with respect to the tabulated BK targets
on a held-out validation set, and should not be interpreted as the intrinsic numerical uncertainty of the BK solver.

Beyond its high accuracy on the validation set, it is important to demonstrate the emulator’s performance on representative test cases. 
In Fig.~\ref{fig:N_vs_r}, we show a direct comparison of the emulated dipole amplitude with the exact BK solution for a parameter set not used in training. 
The excellent agreement across a wide range of $r$ confirms that the emulator reproduces the BK evolution with high fidelity, providing a reliable surrogate for phenomenological applications.

\section{Application to DIS observables}
\label{sec:application_dis}

\begin{figure}[t]
  \centering
  \includegraphics[width=\textwidth]{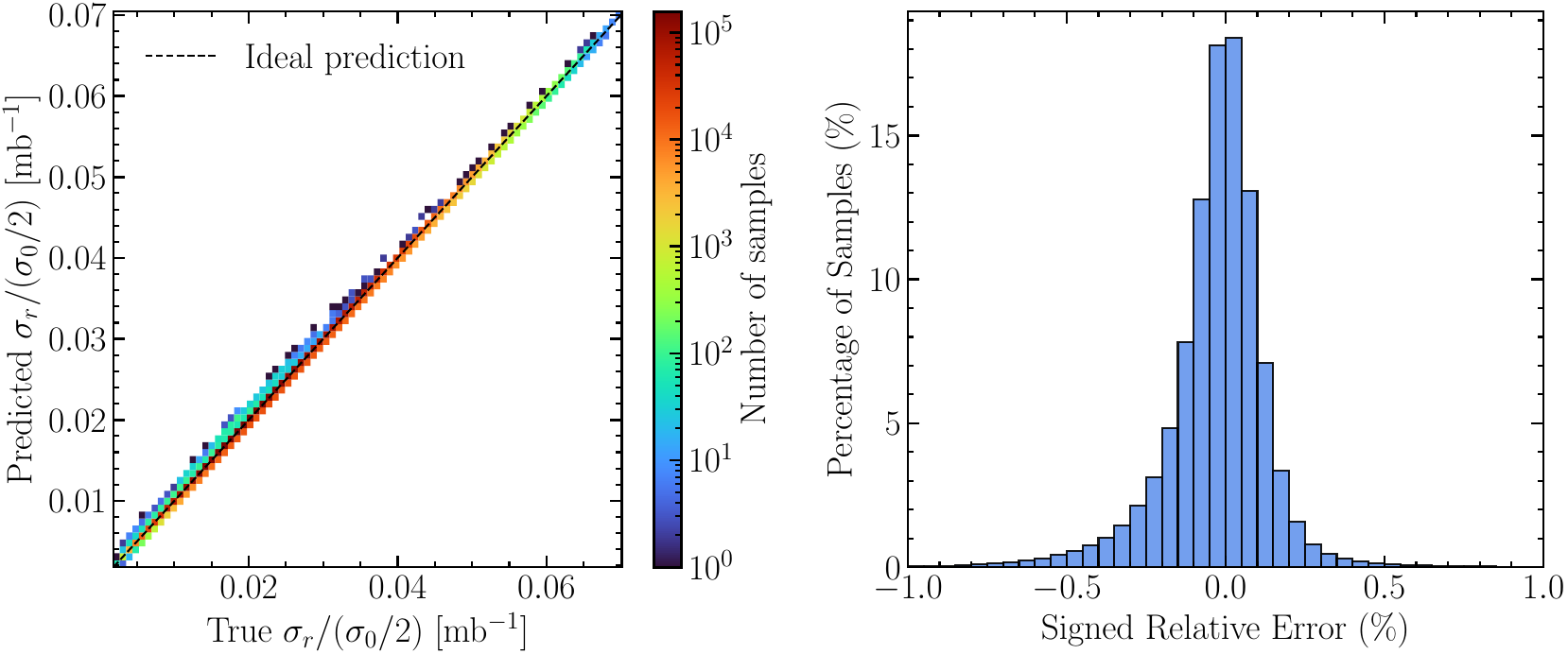}
  \caption{Left: Parity plot comparing the transformer emulator predictions for the DIS reduced cross section $\sigma_r/(\sigma_0/2)$ (y-axis) against the exact dipole model values (x-axis). The dashed line indicates perfect agreement, with mean and median relative errors of $0.115\%$ and $0.073\%$, respectively. Right: Distribution of the signed relative errors, shown within $\pm 1\%$, demonstrating that the overwhelming majority of predictions lie well below the per-mille level. Together these demonstrate the high accuracy of the DIS surrogate model. We find that $99.755\%$ of the validation points fall within the $\pm1\%$ interval.}
  \label{fig:dis_flow}
\end{figure}
With a robust emulator for the BK-evolved dipole amplitude established, we now proceed to its primary application: the calculation of DIS observables for a global analysis of HERA data. This requires convolving the emulated amplitude with the virtual photon wave functions to compute the reduced cross section, $\sigma_r$. 
While these integrals are evaluated offline to build the training set, repeatedly performing the multi-dimensional integrations for each theory evaluation in an iterative analysis across the full HERA kinematics would be computationally expensive.
The surrogate model circumvents this bottleneck by replacing the repeated numerical evaluations with fast predictions, 
thereby enabling efficient exploration of the parameter space. To overcome this final bottleneck, we construct a second, dedicated transformer based emulator. This network is trained to map directly from the six inputs 
$(\ln Q_{s0},\,\gamma,\,e_c,\,\ln C^2,\,\ln x,\,\ln Q^2)$ 
to the normalized reduced cross section $\sigma_r/(\sigma_0/2)$. 
Its architecture follows closely that used for $N(r,x)$, 
and it is trained on the $\sim 4\times 10^6$ values computed from the dipole amplitudes. 

For the DIS cross section emulator, we employ a composite loss function that combines 
mean squared error (MSE) with a Smooth-$L_1$ term~\cite{Terven:2025LossSurvey},
\begin{equation}
    \mathcal{L}_\text{DIS} = 0.8\,\mathrm{MSE}\!\left(\tfrac{\sigma_r^{\rm pred}}{\sigma_0/2}, \tfrac{\sigma_r^{\rm true}}{\sigma_0/2}\right)
                           + 0.2\,\mathrm{SmoothL1}\!\left(\tfrac{\sigma_r^{\rm pred}}{\sigma_0/2}, \tfrac{\sigma_r^{\rm true}}{\sigma_0/2}\right),
\end{equation}
with the Smooth-$L_1$ evaluated at $\beta=0.1$.  
While the MSE provides the primary measure of accuracy, 
we found in practice that adding a small Smooth-$L_1$ component helps stabilize training 
across the wide dynamic range of $\sigma_r/(\sigma_0/2)$ 
by reducing the impact of occasional large residuals.
The resulting surrogate model for the normalized DIS reduced cross section,
$\sigma_r/(\sigma_0/2)$, achieves very high precision, with a mean relative error of
$0.115\%$ and a median error of $0.073\%$. Figure~\ref{fig:dis_flow} illustrates this
accuracy: the left panel shows a parity plot with near-perfect alignment along the diagonal,
while the right panel presents the distribution of signed relative errors within $\pm 1\%$,
with the bulk of predictions clustered well below the per-mille level.

\begin{figure}[t]
  \centering
  \includegraphics[width=\linewidth]{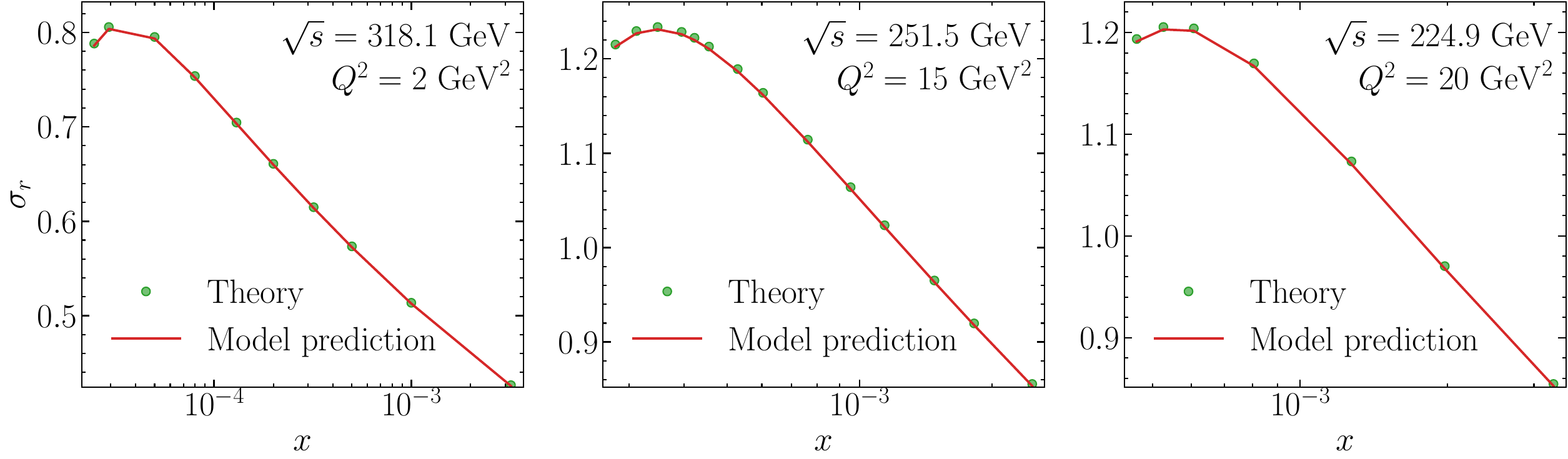}
  \caption{Comparison of the model prediction with the theory cross sections for $e^+p$ DIS at three center-of-mass energies, 
  $\sqrt{s}=318.1$, $251.5$, and $224.9~\mathrm{GeV}$. 
  The green circles denote the theory values, while the red curves show the model prediction evaluated with a representative parameter set outside the training replicas. 
  Each panel is drawn at fixed $Q^2$ ($2$, $15$, $20~\mathrm{GeV}^2$ respectively).}
  \label{fig:band-vs-theory}
\end{figure}

As an intermediate validation step prior to the global fit, 
Fig.~\ref{fig:band-vs-theory} compares the model prediction with the theory cross sections for $e^+p$ DIS at three center-of-mass energies ($\sqrt{s}=318.1,\ 251.5,\ 224.9~\mathrm{GeV}$). 
In each panel the green circles denote the theory values, while the thin red curve shows the prediction evaluated with a representative parameter set outside the training sample.
Each panel is drawn at fixed $Q^2$ values ($2,\ 15,\ 20~\mathrm{GeV}^2$, respectively), with a logarithmic $x$ axis and $\sigma_r$ on the vertical axis. 
The excellent agreement demonstrates that the numerical accuracy of the method is well under control, 
providing a reliable basis for the subsequent phenomenological fits.

This high-speed surrogate enables a full global fit to the combined HERA neutral current $e^+p$ data \cite{H1:2015ubc}. The negligible evaluation cost of the emulator allows for the efficient scanning of millions of points in the parameter space. 
The best-fit parameters $\boldsymbol{\theta}=\bigl(\sigma_0/2,\;Q_{s,0}^2,\;e_c,\;\gamma,\;C^2\bigr)$ are determined by minimizing the chi-squared function
\begin{equation}\label{eq:chisq}
    \chi^2 = \sum_{i=1}^{N_\text{data}} 
    \biggl(\frac{\sigma_{r,i}^{\text{data}} - \sigma_{r,i}^{\text{th}}(\boldsymbol{\theta})}
    {\delta\sigma_{r,i}^{\text{data}}}\biggr)^2,
\end{equation}
where $\sigma_{r,i}^{\text{th}}(\boldsymbol{\theta})$ is the model prediction and $\delta\sigma_{r,i}^{\text{data}}$ denotes the total experimental uncertainty for point~$i$. Statistical uncertainties are estimated using the replica method \cite{Ball:2008by}, in which independent fits are performed on replica data sets generated by adding Gaussian noise to the experimental measurements. This approach has been widely used in the extraction of transverse-momentum-dependent parton distribution functions~\cite{Moos:2025sal,Bacchetta:2024qre,Echevarria:2020hpy}.
In this work we use 200 replicas to obtain the uncertainty estimates. The minimization is performed using the \texttt{Minuit} algorithm from the \texttt{iminuit} package \cite{iminuit}. 

We evaluate the $\chi^2$ using a diagonal approximation for the experimental covariance.
Specifically, the uncertainty in the denominator of Eq.~\eqref{eq:chisq} is taken as a point-by-point total
uncertainty obtained by adding in quadrature the statistical uncertainty, the uncorrelated
systematic uncertainty, and all quoted systematic/procedural contributions provided with the
combined HERA dataset. In the present proof-of-principle LO study we therefore do not introduce
nuisance parameters for correlated systematics. We leave a full correlated treatment (covariance or
nuisance-parameter form) for future work. After applying the kinematic cuts used in this analysis, the fit includes $N_{\rm data}=403$ points
for $x_0=0.01$ and $N_{\rm data}=425$ points for $x_0=0.05$.

\begin{table}[t!]
    \centering
    \begin{adjustbox}{width=\textwidth}
        \begin{tabular}{l c cc cc}
            \toprule
            & & \multicolumn{2}{c}{\textbf{Fit for $\boldsymbol{x_0 = 0.01}$}} & \multicolumn{2}{c}{\textbf{Fit for $\boldsymbol{x_0 = 0.05}$}} \\
            \cmidrule(lr){3-4} \cmidrule(lr){5-6}
            \textbf{Parameter} & \textbf{Prior range} & 4-param & 5-param & 4-param & 5-param \\
            \midrule
            $Q_{s,0}^2$ [GeV$^2$]  
            & $[0.01,\,0.11]$
            & $0.063^{+0.001}_{-0.004}$
            & $0.068^{+0.024}_{-0.015}$
            & $0.025^{+0.0035}_{-0.0025}$
            & $0.045^{+0.0021}_{-0.0017}$ \\
            $e_c$                  
            & $[0.5,\,70.0]$
            & $29.0^{+8.7}_{-3.1}$
            & $18.7^{+20.1}_{-9.6}$
            & $34.0^{+35.3}_{-3.2}$
            & $41.0^{+9.2}_{-2.6}$ \\
            $C^2$                  
            & $[2.0,\,20.0]$
            & $4.45^{+0.99}_{-0.59}$
            & $4.82^{+1.56}_{-2.63}$
            & $17.4^{+2.5}_{-6.1}$
            & $10.2^{+1.6}_{-0.8}$ \\
            $\sigma_0/2$ [mb]     
            & $[12.0,\,20.0]$
            & $14.5^{+0.6}_{-0.4}$
            & $14.8^{+0.9}_{-2.4}$
            & $19.4^{+0.5}_{-2.0}$
            & $16.6^{+0.2}_{-0.4}$ \\
            $\gamma$               
            & $[0.9,\,1.3]$
            & 1.00 (fixed)
            & $1.006^{+0.037}_{-0.019}$
            & 1.00 (fixed)
            & $1.138^{+0.033}_{-0.012}$ \\
            \midrule
            $\chi^2/\mathrm{dof}$ & --- 
            & 0.854 & 0.857 & 1.471 & 1.195 \\
            \bottomrule
        \end{tabular}
    \end{adjustbox}
    \caption{
        Combined fit results for the 4-parameter ($\gamma$ fixed at 1.0) and 5-parameter ($\gamma$ free) models at two starting points $x_0=0.01$ and $x_0=0.05$. 
        Prior ranges reflect the updated bounds in the analysis.
        Entries show the central value with the $95\%$ credible bounds. 
    }
    \label{tab:fitresults_combined}
\end{table}

In our analysis we perform four distinct fits, corresponding to the two choices of the starting point $x_0=0.01$ and $x_0=0.05$, 
and to configurations where the anomalous dimension $\gamma$ is either fixed to $1.0$ or treated as a free parameter with $\gamma=1$ corresponding to the standard MV model.
Our motivation for considering $x_0=0.05$, in addition to the more standard choice $x_0=0.01$, is to assess the validity of the small-$x$ framework for more moderate values of $x$.
The best fit parameters and corresponding $\chi^2$ per degrees of freedom (dof) for all four fit configurations are summarized in Table~\ref{tab:fitresults_combined}.
The overall fit quality is comparable to the recent Bayesian analysis~\cite{Casuga:2023dcf}. 
For our fits with $x_0=0.01$ we obtain $\chi^2/\mathrm{dof}$ values close to unity, in line with the results of Ref.~\cite{Casuga:2023dcf}. 
When the starting point is shifted to $x_0=0.05$, the fit quality deteriorates somewhat, as reflected in the larger $\chi^2/\mathrm{dof}$ values.
Comparing the parameters, we observe that the fitted anomalous dimension $\gamma$ values remain close to 1.0, though the $x_0=0.05$ fit prefers a slightly larger value.
Differences also appear in the extracted $\sigma_0/2$ and $C^2$ parameters, reflecting the sensitivity of the fits to the chosen starting point of the BK evolution.
The smaller $Q_{s,0}^2$ for $x_0=0.05$ is expected, as the saturation scale increases with higher energy (i.e. lower $x$).

\begin{figure}[t]
  \centering
  \includegraphics[width=\linewidth]{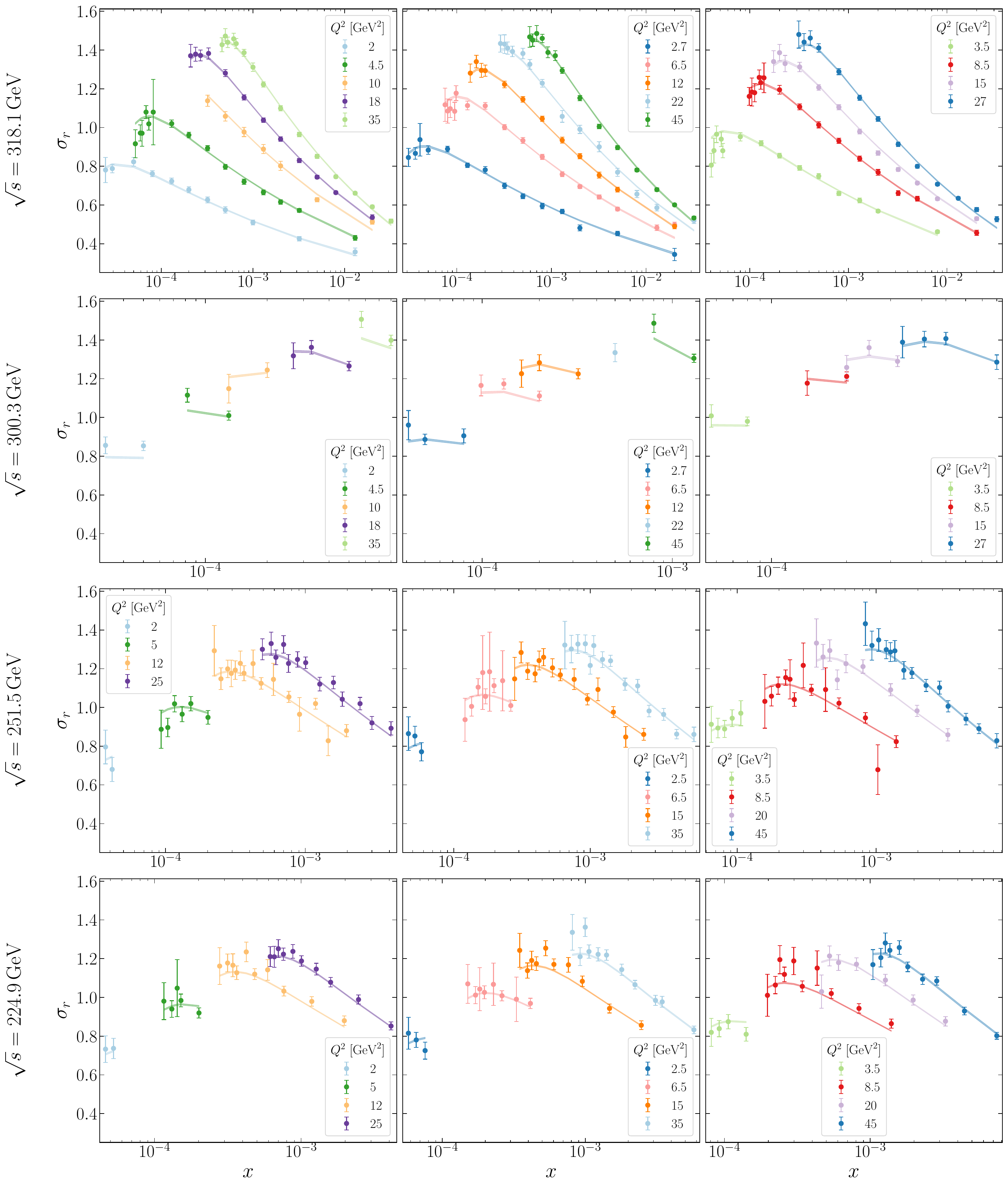}
  \caption{Reduced cross section $\sigma_r(x,Q^2)$ compared with the combined HERA $e^+p$ data at four center-of-mass energies, $\sqrt{s}=318.1$, $300.3$, $251.5$, and $224.9$ GeV. 
  The curves show the result of the five-parameter fit (with $\gamma$ free) performed with the evolution starting point $x_0=0.05$. 
  The shaded bands represent the $2\sigma$ uncertainty of the fitted prediction.}
  \label{fig:dis_fit_x5_5p}
\end{figure}

\begin{figure}[t]
  \centering
  \includegraphics[width=\linewidth]{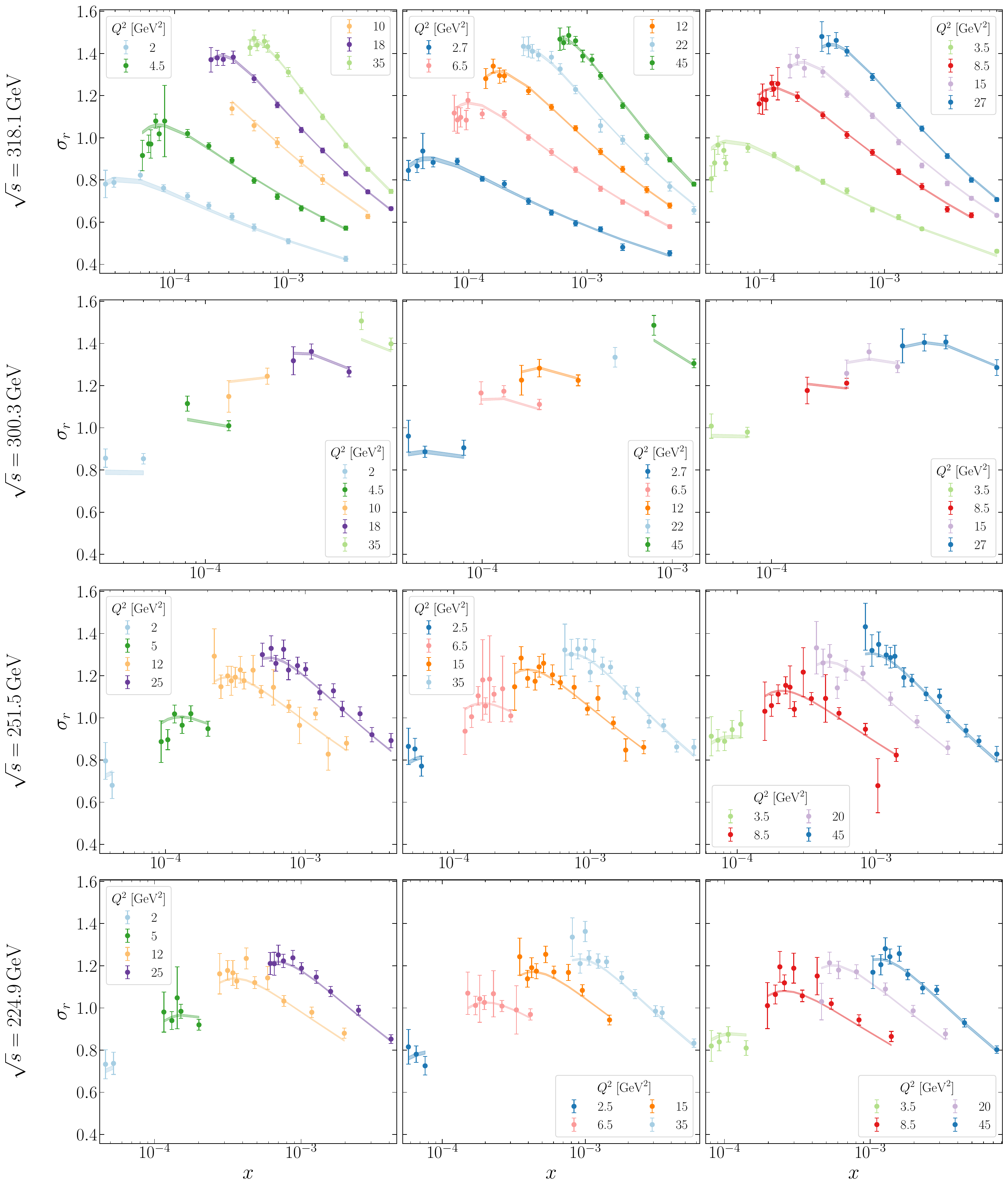}
  \caption{Same as Fig.~\ref{fig:dis_fit_x5_5p}, but with the evolution starting point $x_0=0.01$. 
  The four panels correspond to $\sqrt{s}=318.1$, $300.3$, $251.5$, and $224.9$ GeV. 
  The shaded bands again represent the $2\sigma$ uncertainty of the fitted prediction.}
  \label{fig:dis_fit_x1_5p}
\end{figure}

Figures~\ref{fig:dis_fit_x5_5p} and \ref{fig:dis_fit_x1_5p} present the phenomenological results of this analysis. They provide a direct comparison between our best-fit model and the combined HERA $e^+p$ data for the reduced cross section $\sigma_r$, shown simultaneously across the four experimental center-of-mass energies $\sqrt{s}=318.1$, $300.3$, $251.5$, and $224.9\,\mathrm{GeV}$. The solid curves represent the results of our 5-parameter fit (with $\gamma$ free), using the evolution starting point $x_0=0.05$~(Fig.~\ref{fig:dis_fit_x5_5p}) and $0.01$ (Fig.~\ref{fig:dis_fit_x1_5p}). The shaded bands indicate the $2\sigma$ uncertainty envelope of the fitted prediction, demonstrating an excellent description of the data across the full kinematic range.

\begin{figure}[t]
  \centering
  \includegraphics[width=\linewidth]{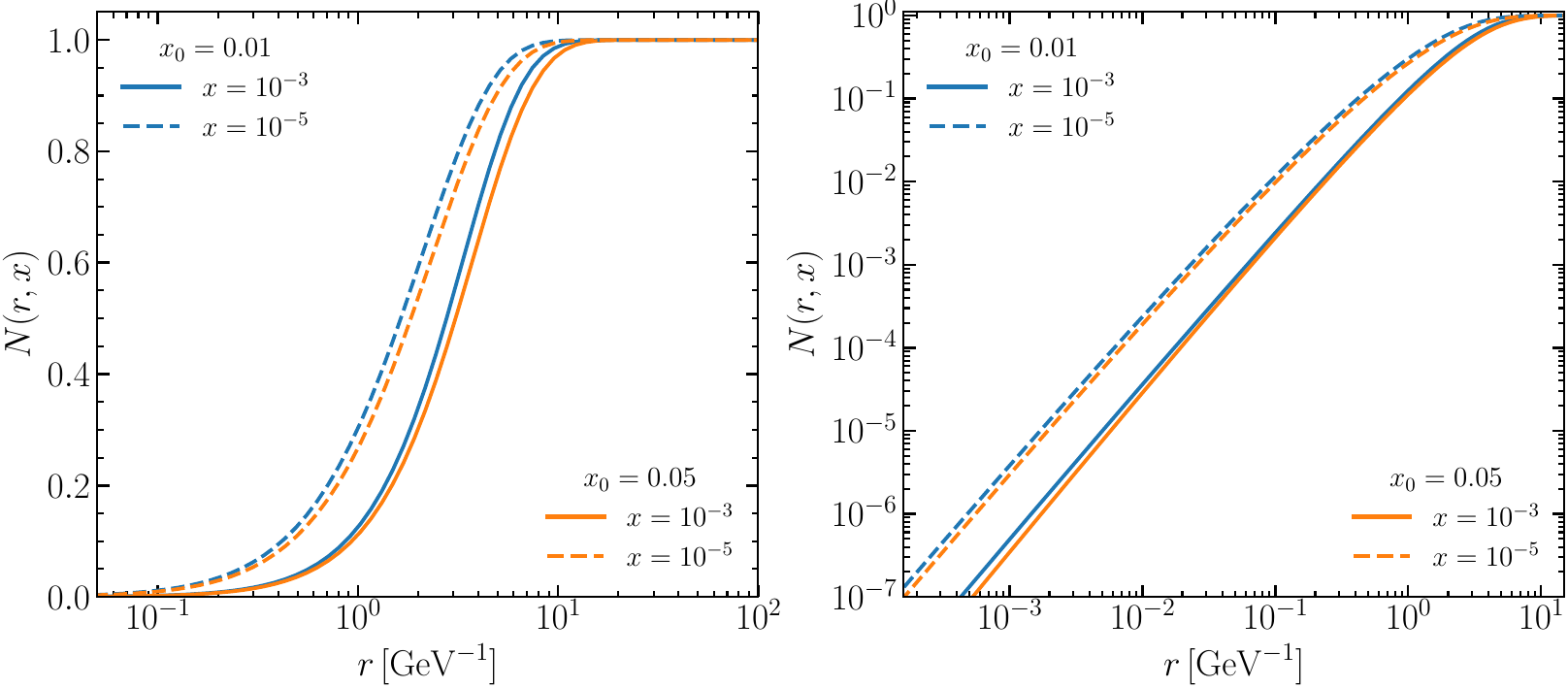}
  \caption{Comparison of the dipole amplitudes $N(r,x)$ from the $x_0=0.01$ 
  (solid lines) and $x_0=0.05$ (dashed lines) from the 5-parameter fits, evaluated at the same values of $x$.
  The left panel shows the results on a linear scale, while the right panel uses a 
  logarithmic scale to emphasize the small-$r$ region. The two sets of curves become 
  nearly identical after the appropriate rapidity shift, with only small residual 
  differences visible around $r\!\sim\!1\ \mathrm{GeV}^{-1}$ and at very small $r$. 
  These remaining discrepancies could be further reduced once the normalization 
  factor $\sigma_0/2$ is included, but here we focus solely on the dipole amplitude itself.
  }
  \label{fig:n-vs-r-x0}
\end{figure}

As a further cross-check, we also examine how the dipole amplitude depends on the choice of the initial rapidity scale. 
In Fig.~\ref{fig:n-vs-r-x0} we compare the fitted dipole amplitudes $N(r,x)$ obtained 
from the $x_0=0.01$ and $x_0=0.05$ fits, evaluated at the same values of $x$. 
Specifically, we compare the results at $x=10^{-3}$ and $x=10^{-5}$ for both initial 
conditions $x_0=0.01$ and $x_0=0.05$. 
The left panel (linear scale) shows that the two curves nearly coincide, with only 
a small residual difference around the transition region $r \!\sim\! 1\ \mathrm{GeV}^{-1}$. 
The right panel (log-log scale) demonstrates that in the small-$r$ region the two amplitudes show a similar behavior, while in the very small-$r$ region (where $N\ll1$) the relative difference can reach the $\mathcal{O}(10\%)$ level. 
This indicates that once the rapidity shift is properly taken into account, the BK 
evolution effectively eliminates the dependence on the arbitrary choice of $x_0$, 
leaving only minor residual differences around the transition region.
The remaining small discrepancy is compensated by a larger value of $\sigma_0/2$ 
for the $x_0=0.05$ fit, such that the produced cross sections are very close to 
each other when away from the saturation region $N \approx 1$.

Finally, we summarize the computational cost of the present workflow to quantify the speed-up enabled by the surrogate strategy.
In our current implementation, constructing the training library requires solving the BK equation for $10^4$ sampled parameter configurations,
with a walltime of about $10$ minutes per configuration on a single CPU core, corresponding to a total of $\sim 1.7\times 10^3$ core-hours.
Training the DIS cross section transformer on a single NVIDIA A100 GPU takes about $40$ hours.
Once trained, the full fit pipeline---including the evaluation of $\sigma_r$ over the HERA kinematics and the subsequent minimization---typically completes in about $2$ minutes.
These benchmarks demonstrate that the emulator removes the dominant bottlenecks of repeated BK evolutions and multi-dimensional integrations in iterative fits,
thereby enabling efficient exploration of the parameter space at negligible per-evaluation cost.

\section{Summary}
\label{sec:discussion}

We have presented a new machine learning framework that, by employing transformer-based emulators, provides high-fidelity representations of both the BK-evolved dipole amplitude and the resulting DIS cross sections. This methodology dramatically reduces the computational cost of small-$x$ phenomenology, enabling parameter space exploration many orders of magnitude faster than direct numerical evolution. Our application of this framework to HERA data confirms that the BK equation, coupled with a three-parameter initial condition, provides an excellent description of inclusive DIS cross sections over a wide kinematic range.

The framework is readily extensible to several crucial areas. Future investigations will include the incorporation of the impact parameter~\cite{Mantysaari:2024zxq,Cepila:2018faq,Cepila:2020xol,Cepila:2023pvh,Cepila:2024qge,Cepila:2025ujv,Bendova:2019psy}, the implementation of next-to-leading order corrections to both the BK kernel~\cite{Balitsky:2007feb,Balitsky:2009xg} and the photon wave functions~\cite{Balitsky:2010ze,Balitsky:2012bs,Beuf:2011xd,Beuf:2016wdz,Beuf:2017bpd,Hanninen:2017ddy,Beuf:2021qqa,Beuf:2021srj,Beuf:2022ndu}, and the application to diffractive DIS measurements both at LO~\cite{Kowalski:2008sa,Munier:2003zb,Marquet:2007nf,Lappi:2023frf} and NLO~\cite{Beuf:2022kyp,Beuf:2024msh,Kaushik:2025roa}. A key advantage of our approach is the decoupling of the computationally intensive evolution from the fitting procedure, which permits the straightforward addition of new parameters or the coupling to other physical processes. This work thus provides a robust and flexible tool, paving the way for the next generation of precision global analyses in small-$x$ physics.
\appendix
\section{Transformer nomenclature used in this work}
\label{app:transformer_nomenclature}

Our emulator adopts an encoder-style Transformer with self-attention~\cite{Vaswani:2017Attention},
layer normalization~\cite{Ba:2016LayerNorm}, and the pre-normalization (Pre-LN) layout commonly used
for stable optimization~\cite{Xiong:2020PreLN}. We summarize here the terminology used in
Sec.~\ref{sec:fitting} and its concrete meaning in our BK-emulator setup.

In this work, one emulator query (and one training sample) corresponds to one input vector
\begin{equation}
\mathbf{x}=\bigl(\ln Q_{s0},\, e_c,\, \gamma,\, \ln C^2,\, \ln (r/{\rm GeV}^{-1}),\, \ln (x_0/x)\bigr),
\end{equation}
which specifies one evaluation point of the target function $N(r,x;\boldsymbol{\theta})$ with
$\boldsymbol{\theta}=(Q_{s0},e_c,\gamma,C^2)$.
The logarithms are used only to improve numerical conditioning; the unit conventions follow those in Sec.~\ref{sec:fitting}.
Each entry $x_i\in\mathbb{R}$ of $\mathbf{x}$ is a scalar feature (a single real number). We map each scalar
to a $d_{\rm model}$-dimensional token (embedding vector) using a shared affine map implemented
as a linear layer applied feature-wise:
\begin{equation}
\mathbf{t}_i = W\,x_i + \mathbf{b}\,, \qquad i=1,\dots,6,
\end{equation}
where $W\in\mathbb{R}^{d_{\rm model}\times 1}$ and $\mathbf{b}\in\mathbb{R}^{d_{\rm model}}$ are trainable
parameters shared across the six features. In other words, the same \texttt{Linear(1,$d_{\rm model}$)}
projection is applied to each scalar feature, producing a length-$6$ sequence of tokens
$\{\mathbf{t}_1,\dots,\mathbf{t}_6\}$ with $\mathbf{t}_i\in\mathbb{R}^{d_{\rm model}}$.

Because self-attention alone does not encode the ordering of tokens, we add a \emph{positional encoding}
to each token before the first encoder layer, following the standard sinusoidal construction introduced
with the Transformer~\cite{Vaswani:2017Attention}. Denoting the positional encoding by $\mathbf{p}_i$,
the input to the encoder stack is $\mathbf{h}_i^{(0)}=\mathbf{t}_i+\mathbf{p}_i$.

\emph{Self-attention} updates each token by allowing it to interact with all other tokens in the sequence.
In scaled dot-product attention~\cite{Vaswani:2017Attention}, one forms queries, keys and values
$Q=HW_Q$, $K=HW_K$, $V=HW_V$ from the matrix $H$ whose rows are the token vectors, and computes
\begin{equation}
{\rm Attention}(Q,K,V)= {\rm softmax}\!\left(\frac{QK^{\mathsf{T}}}{\sqrt{d_k}}\right)V\,.
\end{equation}
Here $H\in\mathbb{R}^{L\times d_{\rm model}}$ stacks the $L=6$ token vectors row-wise; in multi-head attention the projections
are performed separately in each head, with $d_k=d_{\rm model}/n_{\rm head}$.
\emph{Multi-head attention} repeats this operation in parallel with multiple learned projections and
concatenates the results, enabling the model to capture different types of correlations between input
components (here, between parameters and kinematic variables).

A Transformer encoder block consists of a multi-head self-attention sub-layer and a position-wise
feed-forward network (FFN), each wrapped with residual connections and layer normalization.
We use a Pre-LN layout, i.e. the layer normalization is applied inside each residual branch:
\begin{align}
\tilde{H} &= H + {\rm MHA}\bigl({\rm LN}(H)\bigr),\\
H^{+} &= \tilde{H} + {\rm FFN}\bigl({\rm LN}(\tilde{H})\bigr),
\end{align}
where ${\rm LN}$ is layer normalization~\cite{Ba:2016LayerNorm}. The role of Pre-LN versus Post-LN in
optimization stability is discussed in Ref.~\cite{Xiong:2020PreLN}.

After stacking several encoder layers we obtain final token representations $\{\mathbf{h}_i\}$.
We then apply mean pooling $\bar{\mathbf{h}}=\frac{1}{6}\sum_{i=1}^{6}\mathbf{h}_i$ and feed
$\bar{\mathbf{h}}$ to an MLP head to produce a single real-valued \emph{logit}.
Here “logit” denotes the real-valued output before the final sigmoid nonlinearity.
Applying a sigmoid to the logit yields the dipole amplitude $N(r,x)\in(0,1)$, which automatically
enforces the unitarity bound.

\section*{Acknowledgment}
We thank Bjoern Schenke for helpful discussions and Noah Moran for collaboration during the early stage of this project. M.S.G. and D.Y.S. are supported by the National Science Foundations of China under Grant No.~12275052, No.~12147101, and Shanghai Education Committee under Grant No. 24KXZNB04. Z.K. and J.P. are supported National Science Foundation under grant No.~PHY-2515057. This work is also supported by the U.S. Department of Energy, Office of Science, Office of Nuclear Physics, within the framework of the Saturated Glue (SURGE) Topical Theory Collaboration. 

\bibliographystyle{JHEP}
\bibliography{main}

\end{document}